\documentclass[prd,twocolumn,showpacs,floatfix,nofootinbib]{revtex4}
\usepackage{float}
\usepackage{graphicx}
\usepackage{dcolumn}
\usepackage{bm}
\usepackage{ulem}
\graphicspath{{fig/}}
\usepackage[colorlinks=true,linkcolor=blue,citecolor=blue, urlcolor=blue]{hyperref}
\usepackage{adjustbox}

\begin{document}
	
	
	\title{Resolved photoproduction of the $B_c$ meson in electron-proton collisions}

	\author{Na Cai}
	\email{yoloitw@outlook.com}
	\author{Xi-Jie Zhan\footnote{Corresponding author}}
	\email{zhanxj@hbu.edu.cn}
	\affiliation{Department of Physics, Hebei University, Baoding 071002, P.R. China}
	\affiliation{Hebei Key Laboratory of High-precision Computation and Application of Quantum Field Theory, Baoding 071002, P.R. China}
	\affiliation{Hebei Research Center of the Basic Discipline for Computational Physics, Baoding 071002, P.R. China}
	\author{Tai-Fu Feng}
    \email{fengtf@hbu.edu.cn}
	\affiliation{Department of Physics, Hebei University, Baoding 071002, P.R. China}
	\affiliation{Department of Physics, Guangxi University, Nanning, 530004, P.R. China}
	\affiliation{Hebei Key Laboratory of High-precision Computation and Application of Quantum Field Theory, Baoding 071002, P.R. China}
	\affiliation{Hebei Research Center of the Basic Discipline for Computational Physics, Baoding 071002, P.R. China}

	\date{\today}
	
\begin{abstract}
We present a systematic study of $B_c$ meson photoproduction at electron--proton colliders within the framework of nonrelativistic QCD (NRQCD) factorization.
In addition to the dominant direct channel $\gamma+g\to B_c+X$, we include resolved contributions initiated by $g+g$ and $q+\bar q (q=u,d,s)$ subprocesses.
Total cross sections and transverse-momentum distributions are calculated for several collider configurations, including HERA, LHeC, FCC-$ep$, and EIC.
The numerical results show that the direct $\gamma+g$ channel provides the leading contribution over the entire kinematic range.
However, the resolved $g+g$ channel yields a non-negligible correction, reaching the level of $\mathcal{O}(10\%)$ in the low-$p_T$ region  where most events are produced, and it becomes increasingly important at higher collision energies. 
The $q+\bar q$ channel is found to be numerically insignificant.
\end{abstract}


\maketitle

\section{\label{sec:1}Introduction}
The $B_c$ meson is a distinctive heavy-flavored hadron in the Standard Model, consisting of a bottom quark and a charm antiquark (or charge-conjugated state). As the only known meson composed of two heavy quarks of different flavors, the $B_c$ system occupies a unique position between charmonium and bottomonium.
Its spectroscopy, decay properties, and production mechanisms provide sensitive probes of heavy-quark dynamics, nonrelativistic effective theories, and nonperturbative aspects of Quantum Chromodynamics (QCD). 

At high-energy colliders, the production of $B_c$ mesons is governed predominantly by strong interactions. In contrast to quarkonium production, the formation of a $B_c$ meson requires the simultaneous production of both a $b\bar{b}$ and a $c\bar{c}$ pair within a single hard scattering. Consequently, its production rate is suppressed relative to conventional heavy quarkonia, while remaining fully calculable within perturbative QCD. The hard subprocesses involve energy scales set by the heavy-quark masses, where perturbation theory is applicable, whereas the subsequent hadronization into a bound $b\bar{c}$ state probes the transition to the nonperturbative regime. This interplay makes $B_c$ production a sensitive probe of QCD factorization and heavy-quark binding dynamics~\cite{Chang:1992jb,Chang:1994aw}.

Experimentally, the existence of the $B_c$ meson was first firmly established by the CDF Collaboration in 1998 through analyses of proton--antiproton collision data at the Tevatron~\cite{CDF:1998ihx,CDF:1998axz}.
In 2014, the ATLAS Collaboration reported the observation of the radially excited $B_c(2S)$ states~\cite{ATLAS:2014lga}.
These results were subsequently confirmed and refined by the CMS and LHCb Collaborations in 2019 with higher statistical significance and improved mass resolution~\cite{CMS:2019uhm,LHCb:2019bem}.
More recently, further experimental advances have led to the observation of orbitally excited $B_c$ states~\cite{LHCb:2025uce}.
To date, all observations of $B_c$ mesons and their excited states have been obtained exclusively at hadron colliders.

Theoretical investigations of $B_c$ meson production have been pursued across a wide range of collision environments and mechanisms. 
At hadron colliders such as the Tevatron and the LHC, dominant production channels arise from gluon–gluon fusion and heavy-quark fragmentation, calculable within the nonrelativistic QCD (NRQCD) factorization framework at leading and higher orders in $\alpha_s$~\cite{Chang:1992jb, Chang:1994aw, Kolodziej:1995nv, Braaten:1993jn, Berezhnoy:1994ba, Gershtein:1994jw, Chang:1996jt, Berezhnoy:1996ks, Baranov:1997wy, Baranov:1997sg, Cheung:1999ir, Chang:2003cr, Chang:2004bh, Chang:2005bf, Chang:2005wd, Chang:2003cq, Chang:2005hq, Wang:2012ah, Bi:2016vbt, Chen:2018obq, Berezhnoy:2019yei}. 
Beyond hadronic collisions, $B_c$ production has also been studied in the context of high-luminosity electron–positron colliders~\cite{Yang:2011ps, Chen:2013itc, Chen:2013mjb, Sun:2013liv, Chen:2014xka, Sun:2014kva, Sun:2015hhv, Wei:2018xlr,He:2019tig,Yang:2013vba, Yang:2022zpc,Wang:2025rxw}. 
Indirect sources of $B_c$ mesons through decays of heavy particles have been explored~\cite{Chang:1992bb,Chang:2007si,Deng:2010aq,Yang:2010yg,Jiang:2015jma,Jiang:2015pah,Zheng:2019egj,Yang:2019gga,Chen:2020dtu,Zheng:2023atb}, such as the top quark, W boson and Higgs boson, highlighting the role of heavy-particle decay as a complementary production channel at high energies.
An important class of processes at $e^+e^-$ or electron-proton machines involves photoproduction, where initial-state radiation or Weizsäcker–Williams photons interact to produce a $B_c$ meson accompanied by heavy quarks~\cite{Chen:2014xka, Bi:2017nzv,He:2017had,Sun:2020mvl,Chen:2020dtu, Yang:2022yxb}.
The quasi-real photon can interact as a pointlike particle, and at high energies, the photon can also fluctuate into a hadronic state and develop a nontrivial partonic structure. This gives rise to resolved photoproduction processes, in which partons inside the photon participate in the hard scattering. Such resolved contributions are known to play an important role in various quarkonium production processes, particularly in kinematic regions characterized by low transverse momentum and high center-of-mass energy.
Detailed studies at $e^+e^-$ colliders show that both direct $\gamma\gamma\to B_c + b + \bar{c}$ and resolved photon channels can yield substantial production rates under suitable collider configurations~\cite{Zhan:2022etq}.
Previous studies of $B_c$ photoproduction have mostly focused on the direct $\gamma+g$ channel, while the role of resolved photon contributions has received less attention, especially in the kinematic regime relevant for future high-luminosity $ep$ colliders.

In this work, we present a systematic study of $B_c$ photoproduction at electron-proton colliders, including both direct and resolved photon contributions. In addition to the conventional $\gamma+g$ channel, we incorporate resolved processes initiated by $g+g$ and $q+\bar{q}$ ($q=u,d,s$) partonic subprocesses. We perform numerical analyses for several representative collider configurations, including HERA, LHeC, FCC-ep, and EIC, and investigate both total cross sections and differential distributions.
The inclusion of resolved photon contributions enables us to assess the role of the photon's partonic structure in $B_c$ production. In particular, we analyze the energy and transverse-momentum dependence of different production channels and identify the kinematic regions where resolved processes become non-negligible. Such studies are relevant for future experimental programs aiming to explore QCD dynamics in novel regimes and may provide complementary constraints on parton distribution functions in both the proton and the photon.

\section{\label{sec:2}Formulation}

Within the framework of the Weizs\"acker--Williams approximation (WWA), the energy distribution of photons originating from bremsstrahlung emission can be expressed as~\cite{Frixione:1993yw}
\begin{eqnarray}
	f_{\gamma/e}(x) &=& \frac{\alpha}{2\pi}\Bigg[\frac{1 + (1 - x)^2}{x} {\rm log}\frac{Q^2_{\rm max}}{Q^2_{\rm min}} \nonumber\\
	&&+2m_e^2x\left(\frac{1}{Q^2_{\rm max}}
	-\frac{1}{Q^2_{\rm min}}\right)\Bigg],
\end{eqnarray}
where $x = E_{\gamma}/E_{e}$ denotes the fraction of the longitudinal momentum carried by the emitted photon, $\alpha$ is the electromagnetic fine-structure constant, and the kinematic limits are given by $Q^2_{\rm min} = m_e^2 x^2/(1-x)$ and $Q^2_{\rm max} = (E_e \theta_c)^2(1-x) + Q^2_{\rm min}$. The parameter $\theta_c = 32~\mathrm{mrad}$ corresponds to the maximum allowed scattering angle of the outgoing electron, imposed to ensure that the emitted photon remains quasi-real.

\begin{figure*}
	\centering
	\includegraphics[width=.9\textwidth]{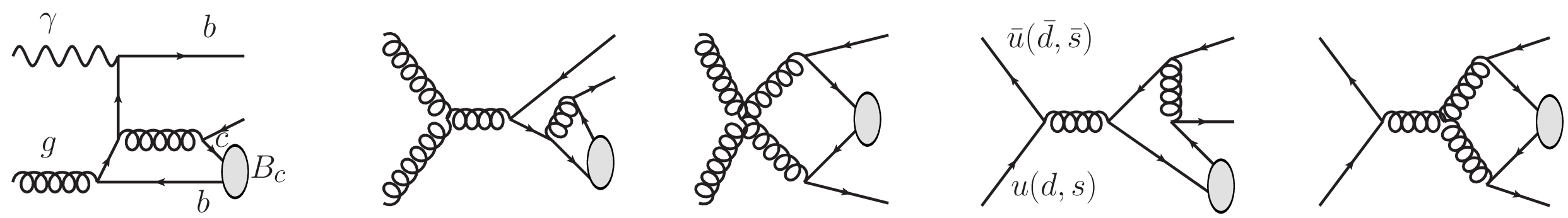}
	\caption{\label{fig:diag} Some typical Feynman diagrams for calculating the partonic cross section $\hat{\sigma}$ of $B_c$ photoproduction at $ep$ collider. The diagrams are drawn by JaxoDraw~\cite{Binosi:2003yf}.}
\end{figure*}

Within the NRQCD factorization formalism, following the approach in Ref.~\cite{Zhan:2022etq} the cross section for $B_c$ photoproduction at an electron--proton collider can be factorized into contributions from the photon flux, the parton distributions in the proton, and the partonic hard-scattering subprocesses,

\begin{eqnarray}
	&&\mathrm{d} \sigma\left(e^{-} + P \rightarrow B_c+X\right)\nonumber\\
	&&= \int \mathrm{d} x_{1}\mathrm{d} x_{2} f_{\gamma / e}\left(x_{1}\right)\sum_{j}f_{j / P}\left(x_{2}\right)
	\sum_{i} \int \mathrm{d} x_{i} f_{i / \gamma}\left(x_{i}\right)\nonumber\\
	&&\times  \sum_{n} \mathrm{~d} \hat{\sigma}(i j \rightarrow c \bar{b}[n]+b+\bar{c})\left\langle \mathcal{O}^{B_c}[n]\right\rangle .
\end{eqnarray}
Here $f_{j/P}$ denotes the parton distribution function (PDF) of parton $j$ inside the proton, while $f_{i/\gamma}$ ($i=\gamma,g,u,d,s$) represents the Gl\"uck--Reya--Schienbein (GRS) distribution function of parton $i$ in the photon~\cite{Gluck:1999ub}. The special case $f_{\gamma/\gamma}(x)=\delta(1-x)$ corresponds to the direct photoproduction contribution. 
The resolved photoproduction channels effectively probe the partonic structure of the quasi-real photon.
The quantity $\mathrm{d}\hat{\sigma}(ij\to c\bar{b}[n]+b+\bar{c})$ is the short-distance partonic cross section, which can be calculated perturbatively within QCD. The intermediate heavy-quark pair $c\bar{b}[n]$ is characterized by the quantum numbers $n$, and the associated long-distance matrix element (LDME) $\langle \mathcal{O}^{B_c}[n]\rangle$ encodes the nonperturbative transition probability for the $c\bar{b}$ pair to hadronize into a physical $B_c$ meson.
In contrast to charmonium and bottomonium systems, where color-octet channels can play an essential role in describing experimental data, the situation for the $B_c$ meson is qualitatively different. The production of a $B_c$ state requires the simultaneous creation of two heavy-quark pairs, which already suppresses the overall production rate. 
As a consequence, fragmentation-type mechanisms that typically enhance color-octet contributions in quarkonium production are less effective in the $B_c$ case.
From the perspective of NRQCD power counting, color-octet contributions are suppressed by powers of the relative velocity $v$ of the heavy quarks inside the bound state. For the $B_c$ system, one expects $v^2 \sim \mathcal{O}(0.1)$, leading to an additional suppression factor of $\mathcal{O}(v^4)$ for the dominant octet channels.
In the present analysis, we restrict ourselves to the leading contribution in the NRQCD velocity expansion, namely the color-singlet channel.
Under this approximation, the corresponding LDME can be evaluated using potential models.

For definiteness, we consider the following partonic subprocesses contributing to three distinct production channels,
\begin{eqnarray}
	\label{eq:channel-1}
	\gamma + g \rightarrow B_c(B^*_c,B_c(2{}^1S_0),B^*_c(2{}^3S_1)) + b + \bar{c},\\
	\label{eq:channel-2}
	g + g \rightarrow B_c(B^*_c,B_c(2{}^1S_0),B^*_c(2{}^3S_1)) + b + \bar{c},\\
	\label{eq:channel-3}
	q + \bar{q} \rightarrow B_c(B^*_c,B_c(2{}^1S_0),B^*_c(2{}^3S_1)) + b + \bar{c},
\end{eqnarray}
where $q=u,d,s$. Representative Feynman diagrams for these subprocesses are shown in Figure~\ref{fig:diag}. Both the analytical derivation and numerical evaluation of the partonic amplitudes are performed using the well-established \textsc{Feynman Diagram Calculation} (FDC) package~\cite{Wang:2004du}. Within this framework, the standard NRQCD projection method~\cite{Bodwin:2002cfe} is employed to extract contributions from specific heavy-quark configurations.

\section{\label{sec:3}Numerical results and discussions}

\begin{table}[b]
	\centering
	\begin{tabular}{|c|ccc|}
		\hline
		colliders & $\sqrt{S}$ & $E_e$ & $E_P$\\
		\hline
		HERA & 319$\mathrm{~GeV}$ & 27.5$\mathrm{~GeV}$ & 920$\mathrm{~GeV}$\\
		LHeC-1 & 1.30$\mathrm{~TeV}$ & 60$\mathrm{~GeV}$ & 7$\mathrm{~TeV}$ \\
		LHeC-2 & 1.98$\mathrm{~TeV}$ & 140$\mathrm{~GeV}$ & 7$\mathrm{~TeV}$ \\
		FCC-$ep$-1 & 7.07$\mathrm{~TeV}$ & 250$\mathrm{~GeV}$ & 50$\mathrm{~TeV}$ \\
		FCC-$ep$-2 & 10.0$\mathrm{~TeV}$ & 500$\mathrm{~GeV}$ & 50$\mathrm{~TeV}$ \\
		EIC-1 & 45$\mathrm{~GeV}$ & 5$\mathrm{~GeV}$ & 100$\mathrm{~GeV}$ \\
		EIC-2 & 140$\mathrm{~GeV}$ & 18$\mathrm{~GeV}$ & 275$\mathrm{~GeV}$\\
		\hline
	\end{tabular}
	\caption{\label{tab:1}The center-of-mass energies($\sqrt{S}$) of representative electron-proton colliders, along with the corresponding energies of the electrons($E_e$) and protons($E_P$).}
\end{table}

\begin{table*}
	\centering
	\begin{adjustbox}{width=\textwidth}
		\begin{tabular}{|c|cccc|}
			\hline
			$colliders$  & $\sigma_{B_c}(\sigma_{\gamma g},\sigma_{gg},\sigma_{q \bar{q}})$&  $\sigma_{B_c(2{}^1S_0)}(\sigma_{\gamma g},\sigma_{gg},\sigma_{q \bar{q}})$ &
			$\sigma_{B^*_c}(\sigma_{\gamma g},\sigma_{gg},\sigma_{q \bar{q}})$ & $\sigma_{B^*_c(2{}^3S_1)}(\sigma_{\gamma g},\sigma_{gg},\sigma_{q \bar{q}})$ \\
			\hline
			HERA(pb) & $0.44(0.43,0.0085,0.0023)$ & $0.27(0.26,0.0051,0.0014)$ & $2.19(2.15,0.020,0.016)$ & $1.31(1.29,0.012,0.0096)$ \\
			LHeC-1(pb) & $3.69(3.51,0.17,0.0088)$ & $2.21(2.10,0.10,0.0052)$ & $16.99(16.51,0.42,0.056)$ & $10.17(9.89,0.25,0.034)$ \\
			LHeC-2(pb) & $6.52(6.12,0.38,0.014)$ & $3.90(3.67,0.23,0.0081)$ & $29.50(28.48,0.94,0.087)$ & $17.66(17.05,0.56,0.052)$ \\
			FCC-$ep$-1(pb) & $23.90(21.49,2.38,0.034)$ & $14.31(12.86,1.42,0.020)$ & $103.15(97.06,5.88,0.21)$ & $61.75(58.11,3.52,0.126)$ \\
			FCC-$ep$-2(pb) & $34.36(30.41,3.90,0.045)$ & $20.57(18.21,2.33,0.027)$ & $147.00(137.07,9.65,0.28)$ & $88.00(82.06,5.78,0.17)$ \\		
			EIC-1(fb) & $0.70(0.67,0.0013,0.030)$ & $0.42(0.400,0.00081,0.018)$ & $3.96(3.71,0.0027,0.25)$ & $2.37(2.22,0.0016,0.15)$ \\	
			EIC-2(fb) & $72.06(70.61,0.68,0.77)$ & $43.14(42.27,0.41,0.46)$ & $378.16(370.97,1.55,5.64)$ & $226.39(222.09,0.93,3.38)$\\						
			\hline
		\end{tabular}
	\end{adjustbox}
	\caption{\label{tab:2}
		The integrated cross sections of the photoproduction of $B_c$ at some typical electron-proton colliders. Each parenthesis contains three numerical values, which correspond respectively to the cross sections of the three production channels, Eq.~(\ref{eq:channel-1}),(\ref{eq:channel-2}) and (\ref{eq:channel-3}).}
\end{table*}

The input parameters in the calculation are taken as follows. The fine structure constant is fixed as $\alpha =1/137$. $m_b=4.8\mathrm{~GeV}$, $m_c=1.5\mathrm{~GeV}$ and $M_{B_c}=m_b+m_c$. The one-loop running strong coupling constant is employed. The renormalization scale is set to be the transverse mass of the $B_c$ meson, $\mu=\sqrt{M^2_{B_c}+p^2_t}$ with $p_t$ being its transverse momentum. The LDMEs $\left\langle \mathcal{O}^{B_c}[n]\right\rangle$ are related to the wave function at the origin, e.g., $\left\langle \mathcal{O}^{B_c}[n]\right\rangle \approx N_c|R_S(0)|^2/(2\pi)$, with $|R_{1S}(0)|^2=1.642\mathrm{~GeV}^3$ and $|R_{2S}(0)|^2=0.983\mathrm{~GeV}^3$~\cite{Eichten:1994gt,Eichten:1995ch}.
Several representative electron-proton colliders~\cite{Karshon:2014iya,LHeCStudyGroup:2012zhm,Acar:2016rde,Accardi:2012qut} were selected, and their corresponding center-of-mass energies are listed in Table~\ref{tab:1}.

The integrated cross sections for the photoproduction of $B_c$ mesons at various electron-proton colliders are summarized in Table~\ref{tab:2}. The results are presented for both ground states ($B_c$, $B_c^*$) and excited states ($B_c(2^1S_0)$, $B_c^*(2^3S_1)$), with contributions from three distinct production channels: the direct photoproduction process $\gamma + g \rightarrow B_c + b + \bar{c}$, and the resolved photoproduction processes $g + g \rightarrow B_c + b + \bar{c}$ and $q + \bar{q} \rightarrow B_c + b + \bar{c}$ (where $q = u, d, s$).
Table~\ref{tab:2} reveals a clear energy dependence in the relative importance of these channels:

\begin{itemize}
	\item At HERA ($\sqrt{S} = 319$ GeV), the resolved $gg$ channel accounts for only $\sim 1.9\%$ of the total $B_c$ cross section, while the $q\bar{q}$ channel is negligible ($\sim 0.5\%$). The direct $\gamma g$ process dominates ($\sim 97.6\%$), consistent with expectations at lower collision energies.
	
	\item At LHeC-2 ($\sqrt{S} = 1.98$ TeV), the $gg$ contribution rises to $\sim 5.8\%$, while the $q\bar{q}$ channel remains small ($\sim 0.2\%$). The $\gamma g$ channel still dominates ($\sim 93.9\%$), but the enhanced gluon density at higher energies begins to manifest.
	
	\item Most significantly, at FCC-$ep$-2 ($\sqrt{S} = 10.0$ TeV), the $gg$ contribution reaches $\sim 11.4\%$, highlighting its importance in high-energy regimes. This substantial contribution underscores the necessity of including resolved processes for precision predictions at future colliders like the FCC-$ep$.
\end{itemize}

This trend can be attributed to the enhancement of the gluon density in the photon at small Bjorken-$x$, which becomes increasingly relevant at higher center-of-mass energies.
The $q\bar{q}$ channel remains consistently negligible across all colliders ($< 1\%$), as expected due to the suppression of quark-initiated processes in photoproduction.
This suppression originates from both the smaller quark densities in the photon and the absence of dynamical enhancements present in gluon-initiated processes.

The cross sections for excited states follow similar patterns, with the $B_c^*$ states having larger cross sections due to their spin-triplet configuration. 
Future electron--proton colliders are designed to operate at very high luminosities.
Taking the FCC-$ep$ as a representative example, its projected integrated luminosity can reach the order of $1~\mathrm{ab}^{-1}$. Combined with the total cross sections listed in Table~\ref{tab:2}, this implies that approximately ${\cal O}(10^{8})$ $B_c$ mesons could be produced at the FCC-$ep$. In contrast, the planned luminosity of the EIC is relatively lower.
After further accounting for realistic experimental detection and reconstruction efficiencies, only a limited number of $B_c$ events are expected to be observable at the EIC.
Although the $\gamma+g$ channel provides the dominant contribution to the total cross section, the $g+g$ resolved channel can reach the level of ${\cal O}(10\%)$, indicating that it constitutes a non-negligible component of $B_c$ photoproduction at $ep$ colliders.

\begin{figure*}
	\includegraphics[width=.24\textwidth]{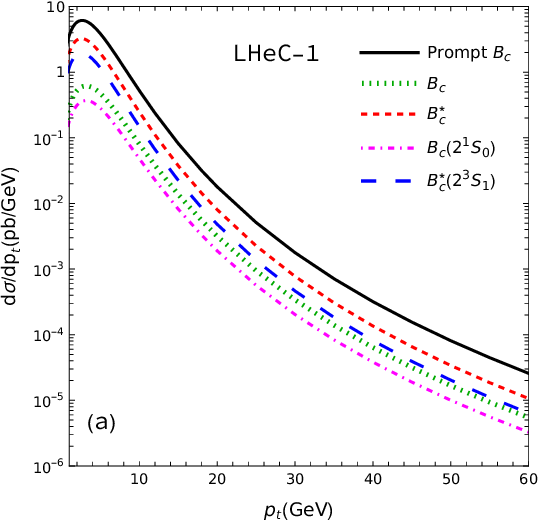}
	\includegraphics[width=.24\textwidth]{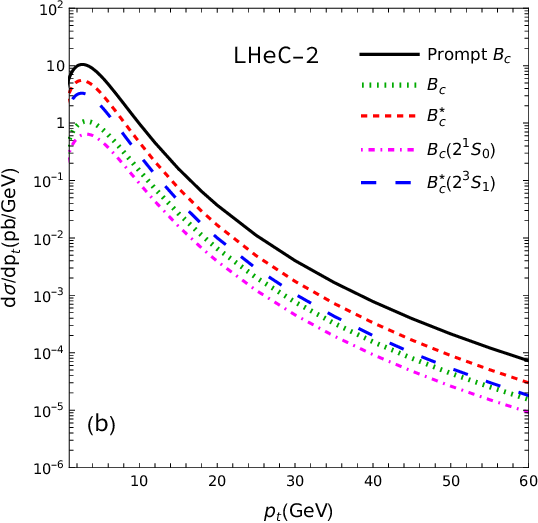}
	\includegraphics[width=.24\textwidth]{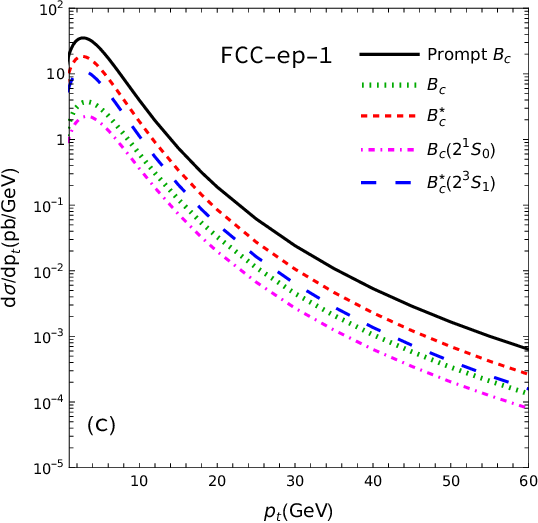}
	\includegraphics[width=.24\textwidth]{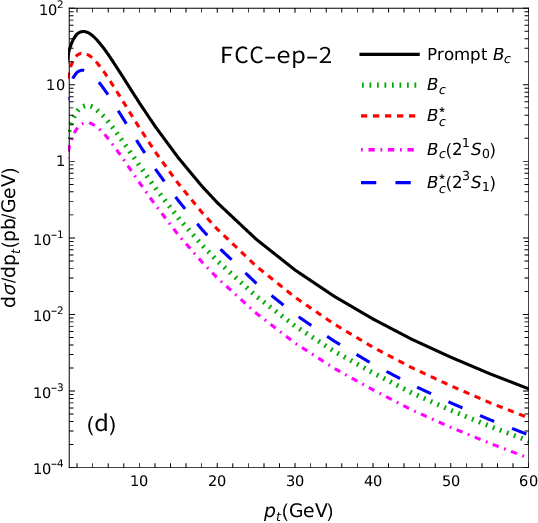}\\
	\includegraphics[width=.24\textwidth]{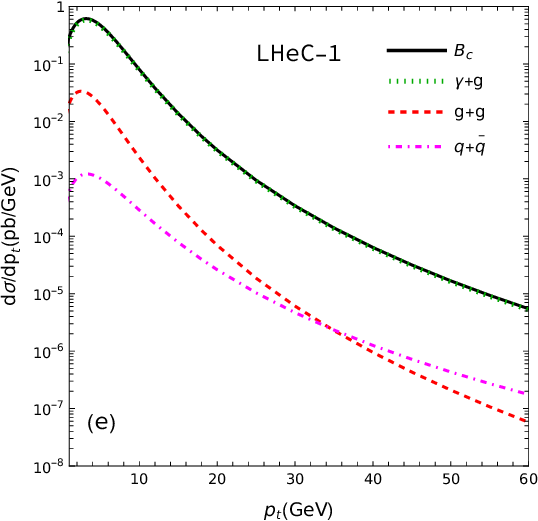}
	\includegraphics[width=.24\textwidth]{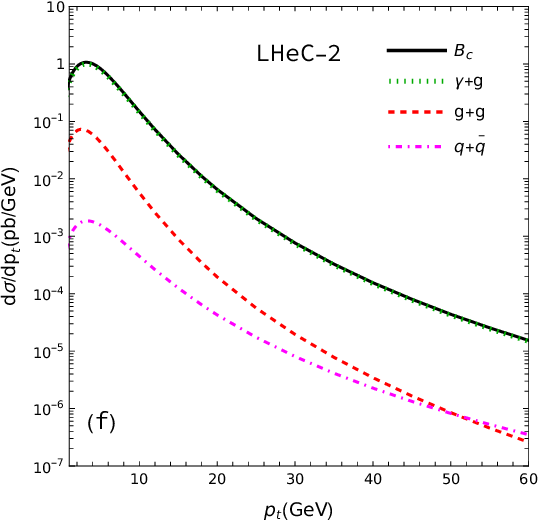}
	\includegraphics[width=.24\textwidth]{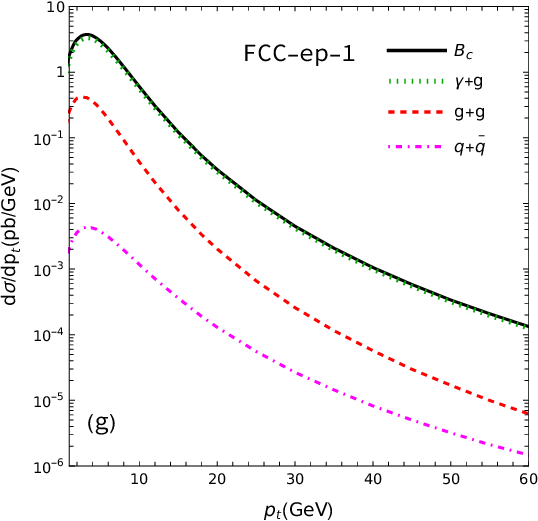}
	\includegraphics[width=.24\textwidth]{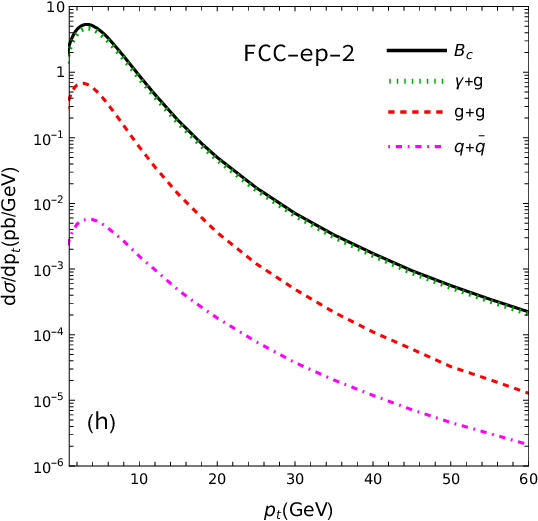}
	\caption{\label{fig:pt} The $p_t$ distributions for $B_c$ photoproduction. (a,b,c,d): $p_t$ distributions of four $B_c$ states and ``prompt $B_c$" means production of the ground $B_c$ after including the feed-down contributions from excited states with $100\%$ decay probability to it.  (e,f,g,h): $p_t$ distributions for the channels in Eqs.~(\ref{eq:channel-1},\ref{eq:channel-2},\ref{eq:channel-3}) of the ground $B_c$ production.}
\end{figure*}

The transverse momentum ($p_t$) distributions for $B_c$ photoproduction provide deeper insights into the kinematic features of different production mechanisms. 
Figure~\ref{fig:pt} presents these distributions for LHeC and FCC-$ep$ configurations.
From Figure~\ref{fig:pt}(a--d), which display the relative contributions to the prompt $B_c$ yield from different intermediate states, one observes a clear dependence on the transverse momentum. 
The distributions for all $B_c$ states ($B_c$, $B_c^*$, $B_c(2^1S_0)$, $B_c^*(2^3S_1)$) exhibit similar shapes, peaking at low $p_t$ and decreasing rapidly with increasing transverse momentum. 
The contribution from the pseudoscalar state $B_c({}^1S_0)$ increases with increasing $p_T$, whereas the contribution from the vector state $B_c^*({}^3S_1)$ shows a mild decrease as $p_T$ grows.

As shown in Figure~\ref{fig:pt}(e--h), the contribution from the $\gamma+g$ channel dominates the photoproduction cross section over the entire $p_T$ region.
This behavior is consistent with the integrated cross-section results presented in Table~\ref{tab:2}.
A closer inspection of the figures reveals that the contribution from the $g+g$ channel decreases with increasing $p_T$, while the $q+\bar{q}$ channel exhibits a relatively harder $p_T$ spectrum, becoming more relevant at larger $p_T$. However, due to its overall suppression, it does not play a significant role in the phenomenology.
As a representative example, at FCC-$ep$-2 the $g+g$ channel contributes about $15.82\%$ of the total cross section at $p_T=1~\mathrm{GeV}$, whereas this fraction is reduced to $5.82\%$ at $p_T=50~\mathrm{GeV}$.
This behavior can be understood from the fact that low-$p_T$ production
probes smaller momentum fractions, where the gluon distributions in both
the proton and the photon are enhanced, thereby increasing the relative
importance of the $g+g$ channel.
In realistic experimental conditions, the majority of events are produced in the low-$p_T$ region.
This further highlights the importance of including the $g+g$ resolved photoproduction channel in theoretical calculations.

\begin{table}
	\centering
	\begin{tabular}{|c|ccc|}
		\hline		
		$m_c(\mathrm{GeV})$ & $1.4$ & $1.5$ & $1.6$\\
		\hline
		$\sigma_{B_c}$ & $8.15(42.64)$ & $6.52(34.36)$ & $5.29(28.08)$ \\
		$\sigma_{B_c(2{}^1S_0)}$ & $4.88(25.53)$ & $3.90(20.57)$ & $3.16(16.81)$ \\			
		$\sigma_{B^*_c}$ & $36.96(183.31)$ & $29.50(147.00)$ & $23.86(120.05)$  \\
		$\sigma_{B^*_c(2{}^3S_1)}$ & $22.13(109.74)$ & $17.66(88.00)$ & $14.28(71.87)$ \\
		\hline
	\end{tabular}
	\caption{\label{tab:uncer-mc}Variations of the integrated cross sections (in unit of pb) by $m_c$ for the photoproduction of $B_c$ under LHeC-2 and FCC-$ep$-2 energy (values in brackets) respectively. Three channels of Eqs.~(\ref{eq:channel-1},\ref{eq:channel-2},\ref{eq:channel-3}) have been summed up.}
\end{table}

\begin{table}
	\centering
	\begin{tabular}{|c|ccc|}
		\hline
		$m_b(\mathrm{GeV})$ & $4.7$ & $4.8$ & $4.9$\\
		\hline
		$\sigma_{B_c}$ & $7.15(37.40)$ & $6.52(34.36)$ & $5.96(31.62)$ \\
		$\sigma_{B_c(2{}^1S_0)}$ & $4.28(22.39)$ & $3.90(20.57)$ & $3.57(18.93)$ \\			
		$\sigma_{B^*_c}$ & $32.30(159.60)$ & $29.50(147.00)$ & $27.00(135.20)$  \\
		$\sigma_{B^*_c(2{}^3S_1)}$ & $19.34(95.54)$ & $17.66(88.00)$ & $16.16(80.94)$ \\
		\hline
	\end{tabular}
	\caption{\label{tab:uncer-mb}Variations of the integrated cross sections (in unit of pb) by $m_b$ for the photoproduction of $B_c$ under LHeC-2 and FCC-$ep$-2 energy (values in brackets) respectively. Three channels of Eqs.~(\ref{eq:channel-1},\ref{eq:channel-2},\ref{eq:channel-3}) have been summed up.}
\end{table}

\begin{table}
	\centering
	\begin{tabular}{|c|ccc|}
		\hline
		${\cal C}$ & $0.5$ & $1.0$ & $2.0$\\
		\hline
		$\sigma_{B_c}$ & $9.73(19.68)$ & $6.52(34.36)$ & $4.58(27.12)$ \\
		$\sigma_{B_c(2{}^1S_0)}$ & $5.82(11.78)$ & $3.90(20.57)$ & $2.74(16.24)$ \\			
		$\sigma_{B^*_c}$ & $43.91(86.59)$ & $29.50(147.00)$ &  $20.78(116.33)$ \\
		$\sigma_{B^*_c(2{}^3S_1)}$ & $26.29(51.84)$ & $17.66(88.00)$ & $12.44(69.64)$ \\
		\hline
	\end{tabular}
	\caption{\label{tab:uncer-mu}Variations of the integrated cross sections (in unit of pb) by $\mu={\cal C}\sqrt{M^2_{B_c}+p^2_t}$ with ${\cal C}=0.5,1,2$, for the photoproduction of $B_c$ under LHeC-2 and FCC-$ep$-2 energy (values in brackets) respectively. Three channels of Eqs.~(\ref{eq:channel-1},\ref{eq:channel-2},\ref{eq:channel-3}) have been summed up.}
\end{table}	

We now turn to the theoretical uncertainties of our predictions.
Tables~\ref{tab:uncer-mc} and \ref{tab:uncer-mb}  show the dependence of the total cross sections on the heavy-quark masses $m_c$ and $m_b$, respectively.
The cross section exhibits a stronger sensitivity to the charm-quark mass than to the bottom-quark mass.
A variation of $m_c$ by $\pm 0.1~\mathrm{GeV}$ around its central value leads to a change of about $20\%$--$30\%$ in the cross section, reflecting the fact that the $c\bar b$ pair is produced close to threshold.
By contrast, varying $m_b$ within the same range induces a more moderate effect, typically at the level of $5\%$--$10\%$.

The uncertainty associated with the choice of renormalization scales is presented in Table~\ref{tab:uncer-mu}.
As expected for a leading-order calculation, the scale dependence constitutes the dominant source of theoretical uncertainty.
Varying the scale from $0.5\,\mu_0$ to $2\,\mu_0$, with $\mu_0=\sqrt{M_{B_c}^2+p_T^2}$, results in variations of the total cross section of order $30\%$ or larger.
This behavior indicates the potential importance of higher-order QCD corrections, which are beyond the scope of the present work.

\section{\label{sec:4}Summary}

In this work, we have performed a systematic analysis of $B_c$ meson photoproduction in electron-proton collisions, incorporating both direct and resolved photon contributions.
The photoproduction of both ground states ($B_c$, $B_c^*$) and excited states ($B_c(2^1S_0)$, $B_c^*(2^3S_1)$) have been investigated. It shows that cross sections of excited states are comparable to, or even larger than, that of the ground state.
Assuming subsequent decays into the ground-state $B_c$, the feed-down contributions significantly enhance the prompt $B_c$ yield and should be included in realistic phenomenological studies.
Our results confirm that the direct $\gamma+g$ channel dominates the $B_c$ production rate over a wide range of collider energies and kinematic regions. Nevertheless, the resolved $g+g$ contribution is found to be non-negligible, particularly in the low transverse-momentum region and at high center-of-mass energies, where it can reach the level of $\mathcal{O}(10\%)$. 
In contrast, the $q+\bar{q}$ channel remains strongly suppressed and has a negligible impact on the phenomenology.
These findings indicate that $B_c$ photoproduction can serve as a complementary probe of the gluon content of the photon, especially in the small-$x$ regime accessible at future high-energy $ep$ colliders.

\begin{acknowledgments}

The work has been supported partly by the National Natural Science Foundation of China (NNSFC) with Grant No. 12305083
and the Natural Science
Foundation of Guangxi Autonomous Region with Grant No. 2022GXNSFDA035068.

\end{acknowledgments}


\end{document}